\documentclass[journal=jcp,manuscript=article]{achemso}
\setkeys{acs}{maxauthors=0}

\usepackage{achemso}
\usepackage{graphics}
\usepackage{amssymb,amsfonts}
\usepackage{graphicx}
\usepackage[table]{xcolor}
\usepackage{caption}
\usepackage{subcaption}
\usepackage{colortbl}
\usepackage{amsmath}
\usepackage{amsopn}
\usepackage{bm}
\usepackage{color}
\usepackage{array}
\usepackage{lscape}
\usepackage{mciteplus}
\usepackage[version=3]{mhchem}
\usepackage{ulem}
\usepackage{listings}
\usepackage{enumerate}
\SectionNumbersOn
\makeatletter
\newcommand*\rel@kern[1]{\kern#1\dimexpr\macc@kerna}
\newcommand*\widebar[1]{%
  \begingroup
  \def\mathaccent##1##2{%
    \rel@kern{0.8}%
    \overline{\rel@kern{-0.8}\macc@nucleus\rel@kern{0.2}}%
    \rel@kern{-0.2}%
  }%
  \macc@depth\@ne
  \let\math@bgroup\@empty \let\math@egroup\macc@set@skewchar
  \mathsurround\z@ \frozen@everymath{\mathgroup\macc@group\relax}%
  \macc@set@skewchar\relax
  \let\mathaccentV\macc@nested@a
  \macc@nested@a\relax111{#1}%
  \endgroup
}
\makeatother
\numberwithin{equation}{subsection}

\author{Janus J. Eriksen}
\email{jeriksen@uni-mainz.de}
\affiliation[Johannes Gutenberg-Universit\"at Mainz]
{Institut f\"ur Physikalische Chemie, Johannes Gutenberg-Universit\"at Mainz, D-55128 Mainz, Germany}
\author{Kasper Kristensen}
\affiliation[Aarhus University]
{qLEAP Center for Theoretical Chemistry, Department of Chemistry, Aarhus University, DK-8000 Aarhus C, Denmark}
\author{Devin A. Matthews}
\affiliation[The University of Texas at Austin]
{The Institute for Computational Engineering and Sciences, The University of Texas at Austin, Austin, Texas 78712, United States}
\author{Poul J{\o}rgensen}
\affiliation[Aarhus University]
{qLEAP Center for Theoretical Chemistry, Department of Chemistry, Aarhus University, DK-8000 Aarhus C, Denmark}
\author{Jeppe Olsen}
\affiliation[Aarhus University]
{qLEAP Center for Theoretical Chemistry, Department of Chemistry, Aarhus University, DK-8000 Aarhus C, Denmark}

\title[TITLE]{Convergence of coupled cluster perturbation theory}

\begin{document}

%
%
\begin{abstract}

The convergence of a recently proposed coupled cluster (CC) family of perturbation series [Eriksen {\it{et al.}}, J. Chem. Phys. {\bf{140}}, 064108 (2014)], in which the energetic difference between two CC models---a low-level parent and a high-level target model---is expanded in orders of the M{\o}ller-Plesset (MP) fluctuation potential, is investigated for four prototypical closed-shell systems (Ne, singlet CH$_2$, distorted HF, and F$^{-}$) in standard and augmented basis sets. In these investigations, energy corrections of the various series have been calculated to high orders and their convergence radii determined by probing for possible front- and back-door intruder states, the existence of which would make the series divergent. In summary, we conclude how it is primarily the choice of target state, and not the choice of parent state, which ultimately governs the convergence behavior of a given series. For example, restricting the target state to, say, triple or quadruple excitations might remove intruders present in series that target the full configuration interaction (FCI) limit, such as the standard MP series. Furthermore, we find that whereas a CC perturbation series might converge within standard correlation consistent basis sets, it may start to diverge whenever these become augmented by diffuse functions, similar to the MP case. However, unlike for the MP case, such potential divergences are not found to invalidate the practical use of the low-order corrections of the CC perturbation series.

\end{abstract}
\newpage
%

%
%
\section{Introduction}\label{intro_section}

In M{\o}ller-Plesset (MP) perturbation theory~\cite{mp2_phys_rev_1934,shavitt_bartlett_cc_book,mest}, the Hartree-Fock (HF) one-determinant wave function is perturbatively corrected for the effect of single (S), double (D), etc., excitations out of the state (up to the level of $N$ excitations, where $N$ is the number of electrons in the system at hand). The target energy for the MP series is thus that of the full configuration interaction (FCI) model. The electronic Hamiltonian is  in MP theory partitioned into the Fock operator, $\hat{f}$ (a zeroth-order part), and the fluctuation potential, $\hat{\Phi}$ (the perturbation), and at lowest (second) order, the resulting MP2 model marks a successful approach to the correlation problem in quantum chemistry, providing a useful, size-extensive doubles correction to the HF energy at a low computational cost. The higher-order corrections of the MP series, however, represent less successful treatments of the electron correlation problem, in part due to the fact that these models offer significantly worse compromises between cost and accuracy than does the MP2 model, and in part because of convergence problems that reflect inherent problems with the actual partitioning of the Hamiltonian. We will now briefly revisit the reason why the MP series might occasionally diverge.

Made possible by an efficient FCI implementation~\cite{olsen_fci_jcp_1988,olsen_fci_cpl_1990}, a study appearing some twenty years ago numerically showed how, upon augmenting standard correlation-consistent basis sets of double-$\zeta$ quality by diffuse functions~\cite{dunning_1_orig,*dunning_5_core}, the MP series was at the risk of becoming oscillating and, ultimately, divergent for a few prototypical, single-reference dominated examples~\cite{mp_divergence_olsen_jcp_1996}. These results were in stark contrast to the behavior of the series within non-augmented standard basis sets, for which the convergence behavior of the MP series had previously been found to reflect the dominance of the reference state in the exact solution~\cite{knowles_handy_mp_cpl_1985,handy_knowles_mp_tca_1985} (at least for MP expansions formulated upon restricted HF (RHF) references~\cite{gill_radom_cpl_1986,nobes_handy_knowles_mp_cpl_1987}). Thus, since the convergence behavior of the MP series was found to exhibit an extreme dependence on the choice of one-electron basis set, even for simple examples dominated by a single determinant, for which no {\it{a priori}} reason for expecting divergences seemed to exist, the usefulness of the higher-order models of the MP series was immediately brought into question, for total energies as well as properties derived from these~\cite{mp_divergence_halkier_jcp_1999}.

The numerical findings mentioned above were subsequently complemented by an explicit determination of the {\it{radius of convergence}} for the MP series~\cite{mp_divergence_christiansen_cpl_1996}. In that study, the electronic Schr{\"o}dinger equation was solved for an electronic Hamiltonian expressed as $\hat{H}(z) = \hat{f} + z\hat{\Phi}$, where $z$ is a complex perturbational strength parameter, such that the zeroth-order (HF) and physical (FCI) wave functions were recovered by the values $z = 0$ and $z = 1$, respectively. When regarded in this way, the MP series is recognized as a power series expansion  in $z$ of the FCI eigenvalues. This expansion will then be divergent if the ground state becomes degenerate with some excited state at a {\it{point of degeneracy}}, $z = \xi$, within the complex unit circle defined by $|z| = 1$~\cite{kato_pert_theory_book,schucan_weidenmuller_ann_phys_1972}. In practice, the authors of Ref. \citenum{mp_divergence_christiansen_cpl_1996} replaced the above search for branch points of the ground state energy function by a search for avoided crossings between the ground state and excited states on the real axis only (i.e., they searched for real-valued points of avoided crossings within the interval $z \in [-1;1]$), since such a search is far less complex than a search within the entire unit circle while still enabling the identification of degeneracies. In this terminology, an excited state is denoted a {\it{back-door}} intruder if an avoided crossing is observed in the interval from $-1<z<0$, while it is denoted a {\it{front-door}} intruder if the crossing falls within the interval from $0<z<1$. A few years later, the analysis in Ref. \citenum{mp_divergence_christiansen_cpl_1996} was supplemented by a simple two-state model capable of explaining the divergence of the MP series~\cite{mp_divergence_olsen_jcp_2000}. In that study, it was illustrated how divergences are bound to occur whenever a basis set is sufficiently flexible to give a reasonable description of highly excited and diffuse back-door intruders, which couple only weakly to the ground state (and, as such, are nearly invisible in the energy spectrum at either $z = 0$ $\lor$ $z= 1$), leading the authors to state that {\it{divergence is the rule rather than exception in MP theory, which converges only in small basis sets}}. This point has since then been repeatedly stressed and extended in related studies~\cite{mp_divergence_larsen_jcp_2000,stillinger_jcp_2000,leininger2000mo,olsen2000convergence,sergeev_jcp_2005,sergeev_jcp_2006,herman_ijqc_2009}.\\

\begin{figure}[htbp]
        \centering
        \includegraphics[scale=0.40]{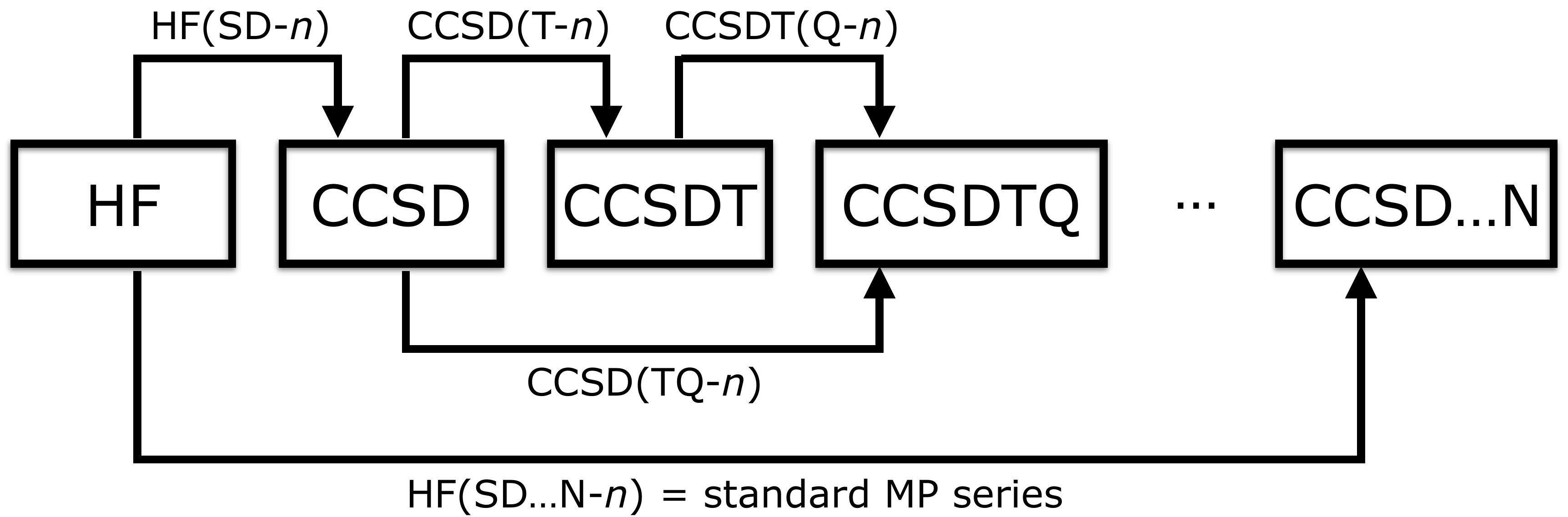}
   \caption{Schematic representation of the HF([$m_{\text{Q}}$]--$n$) and CC[$m_{\text{P}}$]([$m_{\text{Q}}$]--$n$) families of series.}
   \label{parent_target_state_overview_figure}
\end{figure}
We have recently presented a range of perturbation series~\cite{ccsd_pert_theory_jcp_2014,eom_cc_pert_theory_jcp_2014,e_ccsd_tn_jcp_2016} that are all based on a coupled cluster~\cite{cizek_1,*cizek_2,*paldus_cizek_shavitt} (CC) wave function rather than an HF wave function as the zeroth order state, while maintaining the MP partitioning of the Hamiltonian. In the present study, we will extend the type of investigation outlined for the MP series above to these perturbative CC series as well as truncated MP series. Our investigation is centered around the bivariational CC perturbation series introduced in Ref.~\citenum{ccsd_pert_theory_jcp_2014}, while a few results for the recently introduced energy-based CC perturbation series~\cite{e_ccsd_tn_jcp_2016} will also be presented for comparison. For the sake of brevity, we will introduce a general notation, CC[$m$], for the CC model resulting from a truncation of the cluster operator at a specific level, $m$ (i.e., CC[$2$] is the CC singles and doubles~\cite{ccsd_paper_1_jcp_1982} (CCSD) model, CC[$3$] is the CC singles, doubles, and triples~\cite{ccsdt_paper_1_jcp_1987,*ccsdt_paper_2_cpl_1988} (CCSDT) model, etc.). In particular, we want to compare the convergence radii of {\bf{(i)}} a family of so-called HF([$m_{\text{Q}}$]--$n$) expansions, in which the correlation energy of a target CC[$m_{\text{Q}}$] model is expanded in the fluctuation potential, with {\bf{(ii)}} the bivariational  CC[$m_{\text{P}}$]([$m_{\text{Q}}$]--$n$) expansions~\cite{ccsd_pert_theory_jcp_2014}, which describe the energetic difference between a parent model CC[$m_{\text{P}}$] and a target model CC[$m_{\text{Q}}$], again in orders of the fluctuation potential, cf. \ref{parent_target_state_overview_figure}. Thus, the HF([$m_{\text{Q}}$]--$n$) series are MP-like expansions, which theoretically converge from the HF energy, not towards the exact FCI energy, but rather towards the energy of a CC target model, e.g., the target energy of the HF(SDTQ--$n$) series is the CCSDTQ~\cite{ccsdtq_paper_1_jcp_1991,*ccsdtq_paper_2_jcp_1992} (CC singles, doubles, triples, and quadruples) energy. These series are thus equivalent to the standard MP series whenever the latter is restricted to at most $[m_{\text{Q}}]$ excitations. As examples of the second type, the CCSD(T--$n$), CCSD(TQ--$n$), and CCSDT(Q--$n$) series have all recently been proposed~\cite{triples_pert_theory_jcp_2015,open_shell_triples_jcp_2016,quadruples_pert_theory_jcp_2015,open_shell_quadruples_jcp_2016}, forming order expansions that converge from either the CCSD or CCSDT energies onto the CCSDT or CCSDTQ energies, subject to the same partitioning of the electronic Hamiltonian as that used in MP theory, i.e., $\hat{H} = \hat{f} + \hat{\Phi}$. Through lowest (second and third) orders, these series encompass (or are related to) a number of established perturbation models developed within different theoretical frameworks~\cite{ccsd_2_model_gwaltney_head_gordon_cpl_2000,*ccsd_2_model_gwaltney_head_gordon_jcp_2001,ccsd_pt_models_hirata_jcp_2001,*ccsd_pt_models_hirata_jcp_2004,*ccsd_pt_models_hirata_jcp_2007,piecuch_renormalized_cc_0,*piecuch_renormalized_cc_1,*piecuch_renormalized_cc_new_1,*piecuch_renormalized_cc_new_2,*piecuch_renormalized_cc_new_3,kallay_gauss_1,*kallay_gauss_2}; we here refer the reader to Refs. \citenum{ccsd_pert_theory_jcp_2014}, \citenum{triples_pert_theory_jcp_2015}, and \citenum{quadruples_pert_theory_jcp_2015} for recent reviews and comparisons of the various different triples and quadruples models that exist in the literature.

For both families of expansions (based on either an HF or a CC reference state), we will here probe for possible back- and front-door intruders in the real-valued interval $z \in [-1;1]$ by searching for those $z$-values at which the ground state becomes degenerate with an excited intruder state. We consider four examples, all of which have previously been shown to have slowly convergent or divergent MP series---the Ne atom, the F$^{-}$ anion, and the singlet CH$_2$ and HF molecules at equilibrium and distorted geometries, respectively~\cite{mp_divergence_olsen_jcp_1996,mp_divergence_christiansen_cpl_1996,mp_divergence_olsen_jcp_2000}. Furthermore, numerical results for the CC-based perturbation series for different choices of parent and target models will be reported in order to assess to what degree potential divergences might influence these. In this respect, we wish to distinguish between two types of convergence, namely {\bf{(i)}} formal convergence, i.e., {\it{does a given expansion indeed converge?}}, and {\bf{(ii)}} practical convergence, i.e., {\it{are the lowest-order corrections of a given series physically meaningful?}} Thus, while the formal convergence may be assessed by explicitly determining the radius of convergence, the practical convergence may be assessed simply by inspecting the lowest-order energy corrections. For this purpose, we have implemented arbitrary-order energy corrections for all members of the CC[$m_{\text{P}}$]([$m_{\text{Q}}$]--$n$) family of CC perturbation series.

%
%
\section{Theory}\label{theory_section}

We will now outline the theory behind the intruder state scan and the computation of CC[$m_{\text{P}}$]([$m_{\text{Q}}$]--$n$) energy corrections to arbitrary order. In \ref{theory_formal_subsection}, we discuss how the radii of convergence for the perturbation expansions are defined by the largest perturbation parameter for which the CC Jacobian remains non-singular. In \ref{theory_scan_section}, we use this insight to scrutinize the theoretical origin of possible intruder states from a comparison of the Jacobians for HF- and CC-based perturbation series. Finally, in \ref{theory_order_corrections_section}, we review our implementation of general-order CC[$m_{\text{P}}$]([$m_{\text{Q}}$]--$n$) energy corrections.
%

%
%
\subsection{Criteria for the convergence of CC perturbation theory}\label{theory_formal_subsection}

The original analysis of the convergence of MP perturbation theory discussed in \ref{intro_section} may not directly be applied to the present CC context, as the energies and wave function parameters are now obtained from a non-linear set of equations, rather than as the solutions to a standard Hermitian eigenvalue problem for the Hamiltonian operator
\begin{align}
\hat{H}(z) = \hat{f} + z\hat{\Phi} \ . \label{hamiltonian_scaled}
\end{align}
In CC theory, the energy is in general an analytic algebraic function of the cluster parameters, $\{\textbf{t}\}$, and the perturbation strength, $z$, i.e., $E^{\text{CC}} = E^{\text{CC}}(z,\textbf{t})$. In searching for intruder states, the task boils down to determining the range of $z$ within which the energy may be expressed in terms of an analytic function of $z$, or, equivalently, as a convergent Taylor series in $z$. In turn, this range of $z$ is identical to the one where the cluster amplitudes may be formulated as analytic functions of $z$. In order to actually determine this range, we write the non-linear set of equations that define the cluster amplitudes on the generic form
\begin{align}
\textbf{v}(z,\textbf{t}) = \bf{0} \label{eq_cceqs}
\end{align}
with one equation for each individual cluster amplitude. Assuming that for $z = z_0$, the amplitudes $\{\textbf{t}_0\}$ satisfy the equations $\textbf{v}(z_0,\textbf{t}_0) = \bf{0}$, the implicit function theorem~\cite{mathbook_on_implicit} will ensure that the amplitudes are indeed analytic functions of $z$ in a neighborhood of $z_0$, provided that the CC Jacobian, i.e., the derivative of \ref{eq_cceqs} with respect to $\{\textbf{t}\}$, is non-singular in the point $(z_0,\textbf{t}_0)$. By examining the Jacobian as a function of $z$, we may determine a value $\xi$ as that value of $z$ that has the smallest norm while still giving rise to a singular Jacobian. For $z<|\xi|$, the cluster amplitudes---and hence the energy---is thus analytic, implying that $E^{\text{CC}}(z,\textbf{t})$ may be expanded as a convergent Taylor expansion around $z=0$.

In standard CC response theory, the eigenvalues of the Jacobian are identified as excitation energies, which implicates that a singularity in the Jacobian corresponds to a degeneracy between the ground state and an excited state. Thus, although the convergence criteria for the standard MP series and the present CC perturbation series are obtained in different manners, the radius of convergence is in both cases identified as the lowest value of $|z|$ at which such a degeneracy occurs. Phrased differently, the determination of the radius of convergence requires the determination of a complex variable $z$ that produces a zero-eigenvalue of the Jacobian. As discussed in Ref.\citenum{mp_divergence_olsen_jcp_2000}, the imaginary part of this {\textit{branch point}} is in general small for back-door intruders, whereas it may be large for front-door intruders. In both cases, its approximate location may be determined from a scan over the eigenvalues of the Jacobian for real values of $z$, and on par with the procedure in Refs. \citenum{mp_divergence_christiansen_cpl_1996} and \citenum{mp_divergence_olsen_jcp_2000}, we will therefore only examine the lowest eigenvalue of the Jacobian as a function of real values of $z$. Thus, instead of exact degeneracies, we will probe for avoided crossings where the real value of the lowest Jacobian eigenvalue reaches a minimum. For a back-door intruder, the smallness of the imaginary part of the branch point will often lead to very pronounced avoided crossings with the lowest eigenvalue approaching zero, whereas for front-door intruders, the large imaginary part of the branch point tends to produce less pronounced avoided crossings.

Generally speaking, the CC Jacobian is a non-symmetric matrix. This fact leads to two main differences compared to the standard approach of FCI, for which eigenvalues of a Hermitian matrix are determined (assuming a vanishing imaginary component, $\operatorname{Im}(z) = 0$). First, a non-symmetric matrix needs not have a complete set of eigenvalues and eigenvectors. Thus, whereas the singularity of the Jacobian matrix at the branch point trivially corresponds to a zero eigenvalue, it cannot in general be ensured that this eigenvalue changes into another very small eigenvalue for a value of $z$ infinitesimally close to the branch point. However, we note that such abrupt 'dissolutions' of zero eigenvalues were never encountered in the course of the present study. Second, the eigenvalues of a non-symmetric matrix may be complex. Thus, whereas for the Hermitian case, the occurrence of an intruder state lower in energy than the reference state is a clear sign of a singularity (and hence a branch point), this may not need be the case for a non-Hermitian problem. However, since complex eigenvalues of a real matrix always occur in pairs with identical real parts and opposite imaginary parts, and due to the analyticity of the eigenvalues of a general non-degenerate matrix as functions of its elements, a real eigenvalue may only turn into a complex eigenvalue at a branch point. The immediate consequence of this is the practical convenience of being able to determine branch points in much the same way for CC perturbation theory as was previously realized for MP theory in Refs. \citenum{mp_divergence_christiansen_cpl_1996} and \citenum{mp_divergence_olsen_jcp_2000}.

%
%
\subsection{Scan for intruder state}\label{theory_scan_section}

In formulating either of the HF- or CC-based series discussed in \ref{intro_section} as order expansions in a perturbational strength parameter, $z$, we partition the electronic Hamiltonian according to \ref{hamiltonian_scaled}. Next, we define the amplitude equations (\ref{eq_cceqs}) needed for the evaluation of the energy corrections in the series ($\{\textbf{t}(z)\}$ and $\{\tilde{\textbf{t}}(z)\}$ amplitudes, respectively, for the HF- and CC-based series)
\begin{subequations}
\label{ampl_eqs_collected}
\begin{align}
0 &= \langle \mu_{\text{P}} | \exp{(-\tilde{T}(z))}\hat{H}(z)\exp{(\tilde{T}(z))} - \exp{(-{^{\ast}}T)}\hat{H}(z)\exp{({^{\ast}}T)} | \text{HF} \rangle \label{cc_p_ampl_eq} \\
0 &= \langle \mu_{\text{Q}} | \exp{(-T(z))}\hat{H}(z)\exp{(T(z))} | \text{HF} \rangle \label{hf_cc_q_ampl_eq} \\
0 &= \langle \mu_{\text{Q}} | \exp{(-\tilde{T}(z))}\hat{H}(z)\exp{(\tilde{T}(z))} | \text{HF} \rangle \ . \label{cc_q_ampl_eq}
\end{align}
\end{subequations}
In \ref{ampl_eqs_collected}, the $T(z)$, $\tilde{T}(z)$, and ${^{\ast}}T$ cluster operators are all generically defined as $T = \sum_{i}\sum_{\mu_i}t_{\mu_i}\hat{\tau}_{\mu_i}$ (with $t_{\mu_i}$ and $\hat{\tau}_{\mu_i}$ being the amplitude and excitation operator, respectively, for excitation $\mu_i$ within the manifold at level $i$), and the $\{{^{\ast}}\textbf{t}\}$ amplitudes are those of parent model CC[$m_{\text{P}}$] (for an HF parent state, these trivially vanish). Furthermore, the P- and Q-spaces define the so-called primary and secondary (complementary) excitation manifolds, such that the $\{{^{\ast}}\textbf{t}\}$ amplitudes have components only within the primary space, the $\{\textbf{t}(z)\}$ amplitudes exist in the complementary space only (i.e., the primary space is empty for HF-based series), while the $\{\tilde{\textbf{t}}(z)\}$ amplitudes have components in both of the two subspaces. For example, for the MP-like HF(SDT--$n$) series, the P-space will be empty (as is the case for all HF([$m_{\text{Q}}$]--$n$) series) and the Q-space will be the complete manifold of all single, double, and triple excitations, while for the CCSD(T--$n$) series, the P-space will contain all single and double excitations whereas the Q-space will be restricted to the triples manifold only. A schematic illustration of some of these models is given in \ref{parent_target_state_overview_figure}.

Since we will search for intruder states by probing for avoided crossings between the ground state and an excited state in the interval $z \in [-1;1]$, the degeneracies will show up as zero-valued excitation energies. As mentioned in \ref{theory_formal_subsection}, excitation energies are determined within CC response theory as eigenvalues of the CC Jacobian, which is defined as the derivative of the amplitude equations in \ref{ampl_eqs_collected} with respect to either of the $\{\textbf{t}(z)\}$ or $\{\tilde{\textbf{t}}(z)\}$ sets of amplitudes. The elements of the Jacobians for the HF([$m_{\text{Q}}$]--$n$) and CC[$m_{\text{P}}$]([$m_{\text{Q}}$]--$n$) series are thus given as
\begin{subequations}
\label{jacobians_collected}
\begin{align}
J_{\mu_i,\nu_j}[z,t(z)] &= \langle \mu_i | \exp{(-T(z))}[\hat{H}(z),\hat{\tau}_{\nu_j}]\exp{(T(z))} | \text{HF} \rangle \label{hf_cc_jacobian} \\
\tilde{J}_{\mu_i,\nu_j}[z,\tilde{t}(z)] &= \langle \mu_i | \exp{(-\tilde{T}(z))}[\hat{H}(z),\hat{\tau}_{\nu_j}]\exp{(\tilde{T}(z))} | \text{HF} \rangle \label{e_cc_jacobian}
\end{align}
\end{subequations}
where $i,j$ may refer to any excitation level up to level [$m_{\text{Q}}$]. As is evident upon comparing the two Jacobians for the HF([$m_{\text{Q}}$]--$n$) and CC[$m_{\text{P}}$]([$m_{\text{Q}}$]--$n$) series in \ref{jacobians_collected}, these have the same structure, differing only in the amplitudes, $\{\textbf{t}(z)\}$ and $\{\tilde{\textbf{t}}(z)\}$. Using the Baker-Campbell-Hausdorff expansion~\cite{mest,merzbacher_bch_exp_book}, the difference between the matrix elements $J_{\mu_i,\nu_j}$ and $\tilde{J}_{\mu_i,\nu_j}$ becomes
\begin{align}
J_{\mu_i,\nu_j} - \tilde{J}_{\mu_i,\nu_j} = z\langle \mu_i | [[\hat{\Phi},\hat{\tau}_{\nu_j}],T(z)-\tilde{T}(z)] + \ldots | \text{HF} \rangle \label{jacobian_difference}
\end{align}
which, in general, makes the eigenvalues of the two Jacobians, i.e., the excitation energies, differ if the $\{\textbf{t}(z)\}$ or $\{\tilde{\textbf{t}}(z)\}$ amplitudes themselves differ notably. For $z = 0$ $\land$ $z = 1$, however, we note how the excitation energies obtained from a diagonalization of \ref{hf_cc_jacobian} and \ref{e_cc_jacobian} will be identical; for $z = 0$, both of the $\textbf{J}$ and $\tilde{\bf{J}}$ matrices are diagonal with orbital energy differences along the diagonal, and for $z = 1$, both return the excitation energies of target model CC[$m_{\text{Q}}$]. 

To a first approximation and for a general $z$, the Jacobians in \ref{jacobians_collected} are both dominated by the common contribution
\begin{align}
J^{\text{com}}_{\mu_i,\nu_j}[z] = \langle \mu_i | [\hat{f}+z{\hat{\Phi}},\hat{\tau}_{\nu_j}] | \text{HF} \rangle \label{jacobian_common}
\end{align}
which is independent of either set of amplitudes. In particular, we note from \ref{jacobian_common} how for $z<0$, the unphysical action of the fluctuation potential will be to attract, rather than repel the electrons~\cite{stillinger_jcp_2000,herman_ijqc_2009}. To see what ramifications this effect might have on the resulting excitation energies, we focus on the dominating diagonal elements of $\textbf{J}^{\text{com}}$ in \ref{jacobian_common}, which may be written as
\begin{align}
J^{\text{com}}_{\mu_i,\mu_i}[z] &= \epsilon_{\mu_i} + z \big(\langle \mu_i | \hat{\Phi} | \mu_i \rangle - \langle \text{HF} | \hat{\Phi} | \text{HF} \rangle \big) \nonumber \\
&= (1-z)\epsilon_{\mu_i} + z \big(\langle \mu_i | \hat{H}(1) | \mu_i \rangle - \langle \text{HF} | \hat{H}(1) | \text{HF} \rangle \big) \label{jacobian_common_diag}
\end{align}
where $\epsilon_{\mu_i}$ is the (positive) difference in energy between the virtual and occupied orbitals of excitation $\mu_i$. For the diagonal elements in \ref{jacobian_common_diag}, the second term is the difference in energy between the determinant for the $\mu_i$-th excited state and that for the HF ground state. Thus, $\big(\langle \mu_i | \hat{H}(1) | \mu_i \rangle - \langle \text{HF} | \hat{H}(1) | \text{HF} \rangle \big)$ will always be positive, and it will be large for a system with a very diffuse excited state (high excited state energy) and/or for an electron-rich system with a dense HF ground state (low HF ground state energy). Hence, assuming that the eigenvalues of $\textbf{J}^{\text{com}}$ will be dominated by the diagonal elements in \ref{jacobian_common_diag}, we may expect these to approach zero for general combinations of electron-rich ground states and diffuse excited states at $z<0$, if the difference in electron repulsion energy counterbalances the orbital energy difference, i.e., if the following condition is met
\begin{align}
(1-z)\epsilon_{\mu_i} = -z\big(\langle \mu_i | \hat{H}(1) | \mu_i \rangle - \langle \text{HF} | \hat{H}(1) | \text{HF} \rangle \big) \qquad (z<0) \label{jacobian_common_condition}
\end{align}
where $\epsilon_{\mu_i}$ increases as well when the virtual states involved are located high up in the energy spectrum. We will test this hypothesis in \ref{num_results_section} by calculating excitation energies, not only from the two Jacobians in \ref{jacobians_collected}, but also from the dominant contribution to these, i.e., $\textbf{J}^{\text{com}}$ in \ref{jacobian_common}. In passing, however, we note that the above reasoning is in line with that of Ref. \citenum{mp_divergence_olsen_jcp_2000}, in which the presence of intruder states was also linked to differences in first-order MP corrections for the ground state and diffuse excited states, i.e., $\big(\langle \mu_i | \hat{\Phi} | \mu_i \rangle - \langle \text{HF} | \hat{\Phi} | \text{HF} \rangle \big)$.

%
%
\subsection{Arbitrary-order energy corrections}\label{theory_order_corrections_section}

In calculating energy corrections to arbitrary order, the simplest procedure is to evaluate these using the standard $n+1$ rule of Rayleigh-Schr{\"o}dinger perturbation theory, which allows for the energy correction at order $n+1$ to be calculated from the $n$th-order amplitude corrections. For the MP series, the resulting recursive algorithm, as formulated within an FCI program, has previously been described in, e.g., Refs. \citenum{mp_divergence_olsen_jcp_1996} and \citenum{handy_knowles_mp_tca_1985}, and will not be repeated here. For the CC[$m_{\text{P}}$]([$m_{\text{Q}}$]--$n$) family of perturbation series, on the other hand, which was originally derived using Wigner's $2n+1$ and $2n+2$ rules, arbitrary-order corrections are here computed using the same $n+1$ rule as used for the MP series. Furthermore, we will also report numbers for the so-called E-CC[$m_{\text{P}}$]([$m_{\text{Q}}$]--$n$) family of series~\cite{e_ccsd_tn_jcp_2016}, which too form order expansions of the difference in energy between a parent and a target CC model in orders of the fluctuation potential. For a given choice of CC[$m_{\text{P}}$] and CC[$m_{\text{Q}}$], the E-CC[$m_{\text{P}}$]([$m_{\text{Q}}$]--$n$) and CC[$m_{\text{P}}$]([$m_{\text{Q}}$]--$n$) series can both be derived from a bivariational energy Lagrangian for target model CC[$m_{\text{Q}}$], by using information exclusively of the right-hand CC state (E-CC[$m_{\text{P}}$]([$m_{\text{Q}}$]--$n$)) or a combination of information on the right- and left-hand ($\Lambda$) states (CC[$m_{\text{P}}$]([$m_{\text{Q}}$]--$n$)) of the parent CC[$m_{\text{P}}$] model. 

For the physical system ($z=1$), the $n$th-order amplitude corrections, from which energy corrections to order $n+1$ may be evaluated in either of the two series, are given as~\cite{e_ccsd_tn_jcp_2016}
\begin{subequations}
\label{cc_n_order_ampl_eqs}
\begin{align}
\delta t_{\mu_\text{P}}^{(n)} &= - \epsilon_{\mu_{\text{P}}}^{-1} \big(\langle \mu_{\text{P}} | [\hat{\Phi}^{{^{\ast}}\hat{T}},\delta\hat{T}] + \tfrac{1}{2} [[\hat{\Phi}^{{^{\ast}}\hat{T}},\delta\hat{T}],\delta\hat{T}] + \ldots | \text{HF} \rangle \big)^{(n)} \label{cc_n_order_ampl_p_space} \\
\delta t_{\mu_{\text{Q}}}^{(n)} &= - \epsilon_{\mu_{\text{Q}}}^{-1} \big(\langle \mu_{\text{Q}} | \hat{\Phi}^{{^{\ast}}\hat{T}} + [\hat{\Phi}^{{^{\ast}}\hat{T}},\delta\hat{T}] + \tfrac{1}{2} [[\hat{\Phi}^{{^{\ast}}\hat{T}},\delta\hat{T}],\delta\hat{T}] + \ldots | \text{HF} \rangle \big)^{(n)} \ . \label{cc_n_order_ampl_q_space}
\end{align}
\end{subequations}
where $\hat{\Phi}^{{^{\ast}}\hat{T}} = \exp{(-{^{\ast}}T)}\hat{\Phi}\exp{({^{\ast}}T)}$ is the CC[$m_{\text{P}}$] similarity-transformed fluctuation potential. For the sake of notational brevity, we have partitioned the correction amplitudes in \ref{cc_n_order_ampl_eqs} into two components, i.e., the corrections to the parent state amplitudes, $\{{^{\ast}}\textbf{t}\}$, in the P-space ($[\text{P}]$) and the full amplitudes in the Q-space ($[\text{Q}]$)
\begin{align}
\delta\hat{T} = \sum_{i\in[\text{P}]}\delta\hat{T}_{i} + \sum_{j\in[\text{Q}]}\delta\hat{T}_{j} = \delta\hat{T}_{\text{P}} + \delta\hat{T}_{\text{Q}} \ .
\end{align}
In the CC[$m_{\text{P}}$]([$m_{\text{Q}}$]--$n$) and E-CC[$m_{\text{P}}$]([$m_{\text{Q}}$]--$n$) series, the difference in energy between the CC[$m_{\text{P}}$] and CC[$m_{\text{Q}}$] states are next expanded in terms of the $E$ and $\bar{E}$ corrections, respectively, according to the following relations
\begin{subequations}
\label{cc_and_e_cc_series}
\begin{align}
E^{\text{CC}[m_{\text{Q}}]} &= E^{\text{CC}[m_{\text{P}}]} + \sum_{n=2}^{\infty}E^{(n)} \qquad (\text{CC}[m_{\text{P}}]([m_{\text{Q}}]\text{--}n)) \label{cc_series_1} \\
E^{\text{CC}[m_{\text{Q}}]} &= E^{\text{CC}[m_{\text{P}}]} + \sum_{n=3}^{\infty}\bar{E}^{(n)} \qquad (\text{E-CC}[m_{\text{P}}]([m_{\text{Q}}]\text{--}n)) \label{e_cc_series_1}
\end{align}
\end{subequations}
where the $n$th-order energy corrections are defined as
\begin{subequations}
\label{cc_and_e_cc_series_2}
\begin{align}
E^{(n)} = \langle {^{\ast}}\lambda | [\hat{\Phi}^{{^{\ast}}\hat{T}},\delta\hat{T}_{\text{Q}}^{(n-1)}] | \text{HF} \rangle + \tfrac{1}{2}\sum_{m=1}^{n-2} \langle {^{\ast}}\Lambda | [[\hat{\Phi}^{{^{\ast}}\hat{T}},\delta\hat{T}^{(m)}],\delta\hat{T}^{(n-m-1)}] + \ldots | \text{HF} \rangle \label{cc_series_2} \\
\bar{E}^{(n)} = \langle \text{HF} | [\hat{\Phi}^{{^{\ast}}\hat{T}},\delta\hat{T}_{\text{P}}^{(n-1)}] | \text{HF} \rangle + \tfrac{1}{2}\sum_{m=2}^{n-3} \langle \text{HF} | [[\hat{\Phi}^{{^{\ast}}\hat{T}},\delta\hat{T}_{\text{P}}^{(m)}],\delta\hat{T}_{\text{P}}^{(n-m-1)}] | \text{HF} \rangle \ . \label{e_cc_series_2}
\end{align}
\end{subequations}
In \ref{cc_series_2}, $\langle {^{\ast}}\Lambda | = \langle \text{HF} | + \langle {^{\ast}}\lambda |$ where the states $\langle {^{\ast}}\lambda | = \sum_{p\in[\text{P}]}\sum_{\nu_{p}}{^{\ast}}\lambda_{\nu_{p}}\langle \text{HF} | \hat{\tau}^{\dagger}_{\nu_{p}}$ are spanned in terms of CC[$m_{\text{P}}$] multipliers, and in \ref{e_cc_series_2}, the summation range for the second contribution is limited by the fact that amplitude corrections within the P-space vanish at first order, cf. \ref{cc_n_order_ampl_p_space}.

Finally, we note that the expansion of the cluster amplitudes in orders of the MP fluctuation potential is the same in the E-CC[$m_{\text{P}}$]([$m_{\text{Q}}$]--$n$) and CC[$m_{\text{P}}$]([$m_{\text{Q}}$]--$n$) series, see \ref{cc_n_order_ampl_eqs}, and the two series therefore share the same Jacobian in \ref{e_cc_jacobian}. 
Hence, the E-CC[$m_{\text{P}}$]([$m_{\text{Q}}$]--$n$) and CC[$m_{\text{P}}$]([$m_{\text{Q}}$]--$n$) series have identical convergence radii, albeit significantly different rates of convergence, as exemplified by the comparison of the E-CCSD(T--$n$) and CCSD(T--$n$) triples series in Ref. \citenum{e_ccsd_tn_jcp_2016}. This point will also be highlighted by the numerical results to follow.

%
%
\section{Numerical results}\label{num_results_section}
\begin{table}[H]
      \caption{Information on the convergence behavior [convergent (con.) / divergent (div.)] for the HF([$m_{\text{Q}}$]--$n$) and CCSD([$m_{\text{Q}}$]--$n$) families of series.
      The aug-cc-pVDZ basis was used for Ne and F$^{-}$, while the cc-pVDZ basis was used for HF (dist.) and CH$_{2}$ ($^1\text{A}_1$).
      }
\label{convergence_table}
\small
{\begin{tabular}{r|l||r|r||r|r||r|r||r|r}
\hline\hline
\multicolumn{1}{c|}{Section} & \multicolumn{1}{c||}{System} & \multicolumn{2}{c||}{[$m_{\text{Q}}$] = 3} & \multicolumn{2}{c||}{[$m_{\text{Q}}$] = 4} & \multicolumn{2}{c||}{[$m_{\text{Q}}$] = 6} & \multicolumn{2}{c}{[$m_{\text{Q}}$] = 8} \\
\cline{3-10}
\multicolumn{1}{c|}{} & \multicolumn{1}{c||}{} & \multicolumn{1}{c|}{HF} & \multicolumn{1}{c||}{CCSD} & \multicolumn{1}{c|}{HF} & \multicolumn{1}{c||}{CCSD} & \multicolumn{1}{c|}{HF} & \multicolumn{1}{c||}{CCSD} & \multicolumn{1}{c|}{HF} & \multicolumn{1}{c}{CCSD} \\
\hline\hline
\multicolumn{1}{r|}{\ref{neon_subsection}} & \multicolumn{1}{l||}{Ne} & con. & con. & con. & con. & {\bf{div.}} & {\bf{div.}} & {\bf{div.}} & {\bf{div.}} \\
\hline
\multicolumn{1}{r|}{\ref{hf_dist_subsection}} & \multicolumn{1}{l||}{HF (dist.)} & con. & con. & {\bf{div.}} & {\bf{div.}} & {\bf{div.}} & {\bf{div.}} & {\bf{div.}} & {\bf{div.}} \\
\hline
\multicolumn{1}{r|}{\ref{methylene_subsection}} & \multicolumn{1}{l||}{CH$_{2}$ ($^1\text{A}_1$)} & con. & con. & con. & con. & con. & con. & \cellcolor{blue!25} & \cellcolor{blue!25} \\
\hline
\multicolumn{1}{r|}{\ref{f_anion_subsection}} & \multicolumn{1}{l||}{F$^{-}$} & {\bf{div.}} & con. & {\bf{div.}} & {\bf{div.}} & {\bf{div.}} & {\bf{div.}} & {\bf{div.}} & {\bf{div.}} \\
\hline\hline
\end{tabular}} \\
\end{table}
In the present Section, we will report the position of the nearest avoided crossing along the real axis within (or outside) the unit circle, as calculated using either $\textbf{J}$ in \ref{hf_cc_jacobian} for the HF([$m_{\text{Q}}$]--$n$) series or $\tilde{\textbf{J}}$ in \ref{e_cc_jacobian} for the CCSD([$m_{\text{Q}}$]--$n$) series ([$m_{\text{P}}$] = 2). As mentioned in the closing paragraph of \ref{intro_section}, the investigations will be made for four prototypical closed-shell examples, all of which are known to have slowly convergent or even divergent MP series upon moving to higher orders in the perturbation. The four examples are: the Ne atom in the aug-cc-pVDZ basis~\cite{dunning_4_aug} (\ref{neon_subsection}), HF (at a distorted geometry of twice the equilibrium bond length, $r_{\text{e}} = 91.6$ pm) and CH$_2$ ($^1\text{A}_1$) ($r_{\text{e}} = 110.7$ pm, $\angle(\text{HCH}) = 102.0^{\circ}$) in the cc-pVDZ basis~\cite{dunning_1_orig} (\ref{hf_dist_subsection} and \ref{methylene_subsection}, respectively), followed by the F$^{-}$ anion in the aug-cc-pVDZ basis (\ref{f_anion_subsection}). The frozen-core FCI level is [$m_{\text{Q}}$] = 6 for CH$_2$ and [$m_{\text{Q}}$] = 8 for Ne, F$^{-}$, and HF. The main results are summarized in \ref{convergence_table}, which reports whether a given series is convergent (con.) or divergent (div.) for a given truncation level, [$m_{\text{Q}}$]. Furthermore, we will report total deviations from frozen-core FCI results for all four systems. The MP and CC[$m_{\text{P}}$]([$m_{\text{Q}}$]--$n$) series start at second order, while for the E-CC[$m_{\text{P}}$]([$m_{\text{Q}}$]--$n$) series, the leading-order correction is of third order, cf. \ref{cc_and_e_cc_series_2}. All of the involved calculations have been performed using the {\sc{lucia}} program~\cite{lucia} with verification of the low-order results for the simplest expansions done using the {\sc{aquarius}} program~\cite{aquarius}. 

%
%
\subsection{The neon atom}\label{neon_subsection}
\begin{figure}[!ht]
        \centering
        \includegraphics[scale=0.70,bb=7 9 577 433]{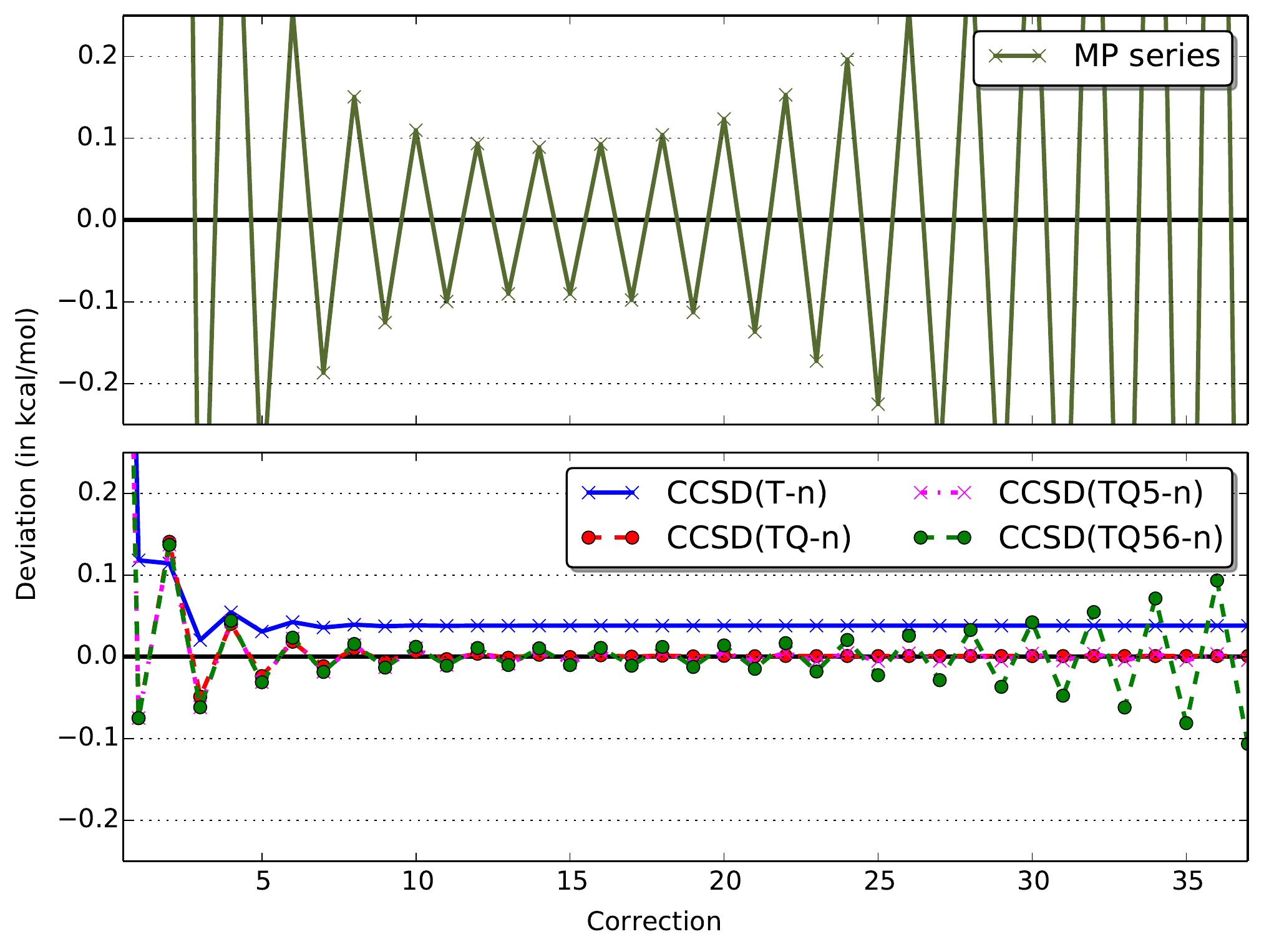}
   \caption{Total deviations (in kcal/mol) of the MP, CCSD(T--$n$), CCSD(TQ--$n$), CCSD(TQ5--$n$), and CCSD(TQ56--$n$) series from the frozen-core/aug-cc-pVDZ FCI correlation energy for Ne.}
   \label{neon_figure}
\end{figure}
In \ref{neon_figure}, we consider the convergence of the neon atom for various perturbation series. In spite of having a substantial weight of the HF determinant in the FCI state (with a resulting HOMO-LUMO energy gap in excess of 30 eV), the MP series for Ne was previously in Ref. \citenum{mp_divergence_olsen_jcp_1996} found to be divergent in the aug-cc-pVDZ basis, due to a weakly coupled back-door intruder with a real component, $\operatorname{Re}(\xi)$, of about $-0.8$, cf. \ref{convergence_table}. If this intruder is decomposed into contributions from individual excitation levels, it is observed (in the MP series, i.e., with [$m_{\text{Q}}$] = 8) to have more than $70\%$ of the weight of its wave function assigned to hextuple and higher-level (seven- and eightfold) excited determinants. Thus, the intruder is clearly an unphysical and diffuse state, which will only appear in augmented basis sets that allow for such continuum states. However, by truncating either of the HF- and CC-based families of series at the level of hextuple excitations, the position of the avoided crossing is observed to move towards a lower $z$-value ($\operatorname{Re}(\xi) \simeq -0.9$ for [$m_{\text{Q}}$] $= 6$), and by further truncating them at the level of pentuple (or lower-level) excitations, the crossing eventually falls outside the unit circle (i.e., $\operatorname{Re}(\xi)<-1.0$ for [$m_{\text{Q}}$] $<6$). Thus, while the MP series does not converge for Ne in the aug-cc-pVDZ basis, for [$m_{\text{Q}}$] $<6$, any of its truncated HF([$m_\text{Q}$]--$n$) series will, as will the corresponding CC[$m_{\text{P}}$]([$m_{\text{Q}}$]--$n$) (and E-CC[$m_{\text{P}}$]([$m_{\text{Q}}$]--$n$), cf. \ref{neon_cc_and_ecc_series_subsubsection}) series. The equivalence between the HF- and CC-based perturbation series in this case is further confirmed by calculations of excitation energies from the $\textbf{J}^{\text{com}}$ Jacobian in \ref{jacobian_common}, for which the crossing is again observed to enter the unit circle upon an inclusion of hextuple excitations in the total excitation manifold, i.e., whenever [$m_{\text{Q}}] \geq 6$.

\subsubsection{The effect of higher-level excitations and basis set diffuseness}\label{neon_divergence_subsubsection}
\begin{figure}[!ht]
        \centering
        \includegraphics[scale=0.63,bb=7 9 558 413]{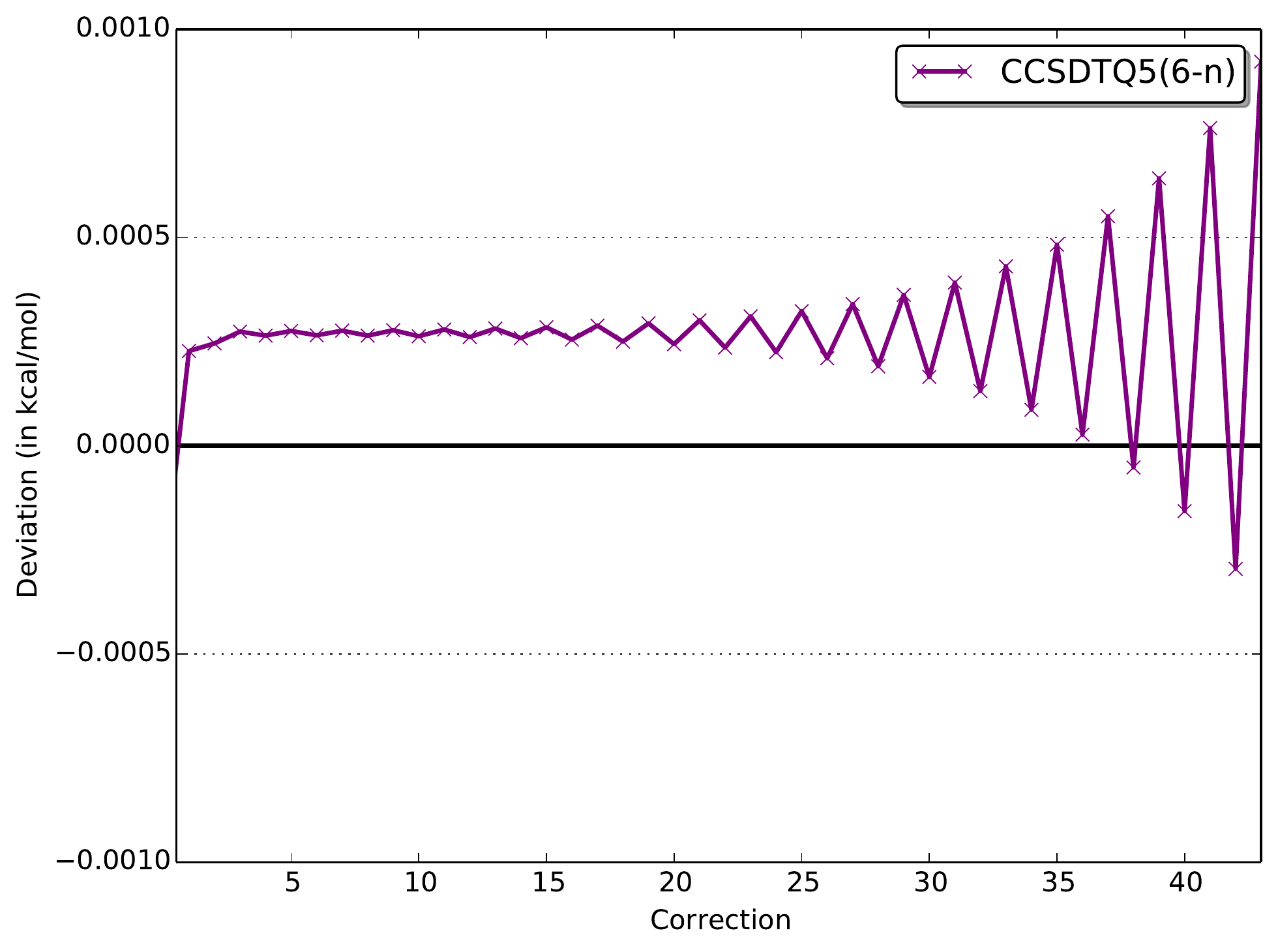}
   \caption{Total deviations (in kcal/mol) of the CCSDTQ5(6--$n$) series from the frozen-core/aug-cc-pVDZ FCI correlation energy for Ne.}
   \label{neon_ccsdtq5_6_figure}
\end{figure}
In the present Section, we want to directly assess the effect of higher-level excitations and basis set diffuseness on the convergence behaviour of the CC[$m_{\text{P}}$]([$m_{\text{Q}}$]--$n$) series. First, in order to quantify that the description of hextuple excitations is indeed at the root of the divergence problems for Ne in the aug-cc-pVDZ basis, we probe for avoided crossings in the CCSDTQ5(6--$n$) series, that is, a series with a CC[5] parent state and a CC[6] target state, for which the difference in energy is a mere $0.0005$ kcal/mol. These results are presented in \ref{neon_ccsdtq5_6_figure}. In summary, an identical crossing with the same intruder state as in the case of the CCSD-based series in \ref{neon_figure} is observed within the unit circle (at $\operatorname{Re}(\xi) \simeq -0.9$), making the series divergent in spite of the minuscule magnitude of the energy difference which the series aims at expanding in orders of the perturbation.

\begin{figure}[!ht]
        \centering
        \includegraphics[scale=0.70,bb=7 9 577 432]{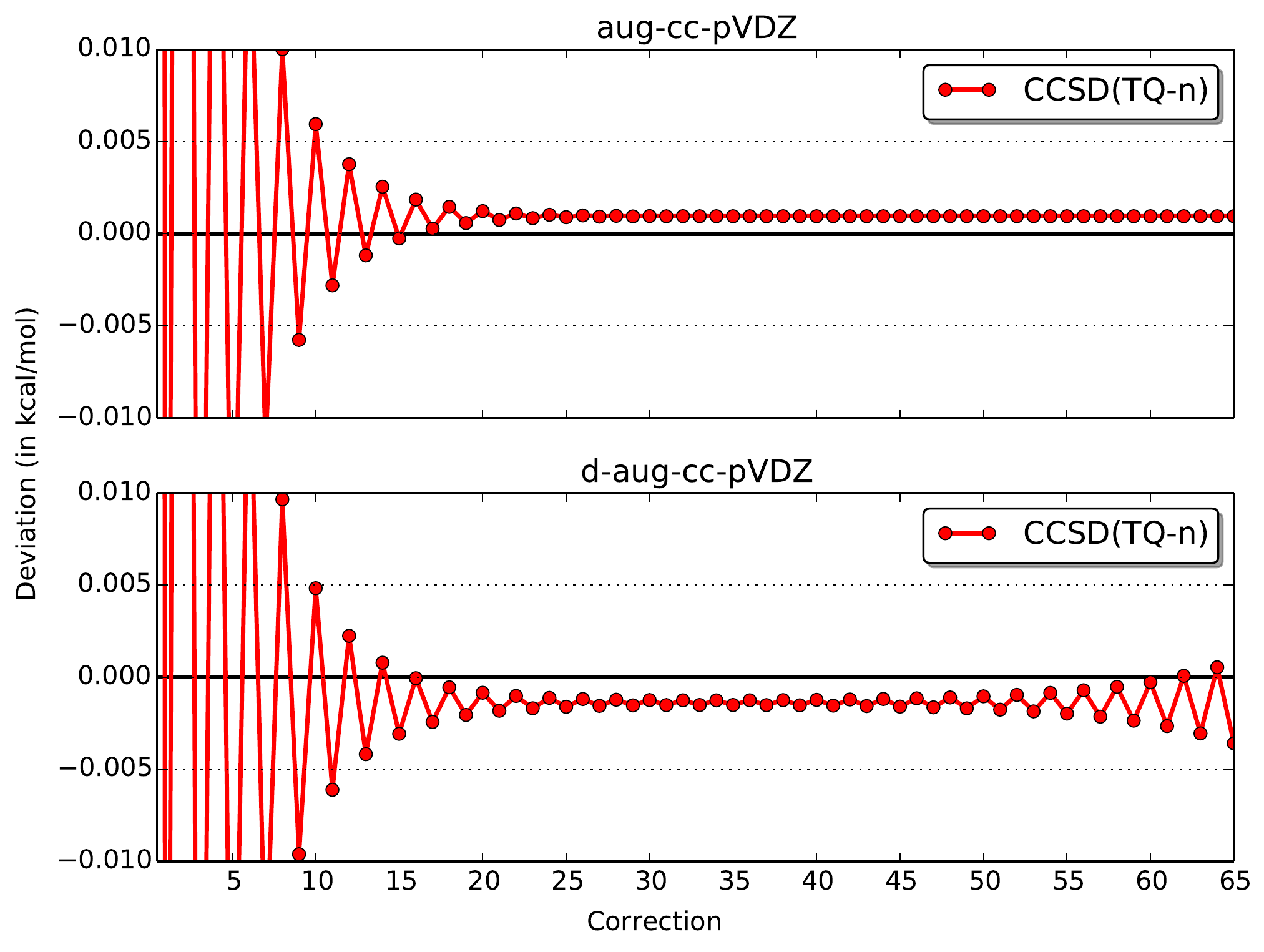}
   \caption{Total deviations (in kcal/mol) of the CCSD(TQ--$n$) series from the frozen-core FCI correlation energy in the aug-cc-pVDZ and d-aug-cc-pVDZ basis sets for Ne.}
   \label{neon_d_aug_ccpvdz_figure}
\end{figure}
Next, we assess whether other intruders but the hextuples-dominated state in the aug-cc-pVDZ basis might enter the unit circle upon an augmentation of the basis by an additional set of diffuse functions. In \ref{neon_d_aug_ccpvdz_figure}, the convergence of the CCSD(TQ--$n$) series is shown in the aug-cc-pVDZ and d-aug-cc-pVDZ basis sets. As is clear, the convergent behaviour of the series in the former of the two basis sets is deteriorated in the latter, as a new intruder dominated by quadruple excitations now enters the unit circle. Thus, in general, a convergent HF- or CC-based series will likely become divergent if the excitation level of the target state is increased and/or the diffuseness of the basis set is increased.

\subsubsection{The CC[$m_{\text{P}}$]([$m_{\text{Q}}$]--$n$) and E-CC[$m_{\text{P}}$]([$m_{\text{Q}}$]--$n$) series}\label{neon_cc_and_ecc_series_subsubsection}
\begin{figure}[!ht]
        \centering
        \includegraphics[scale=0.70,bb=7 9 577 433]{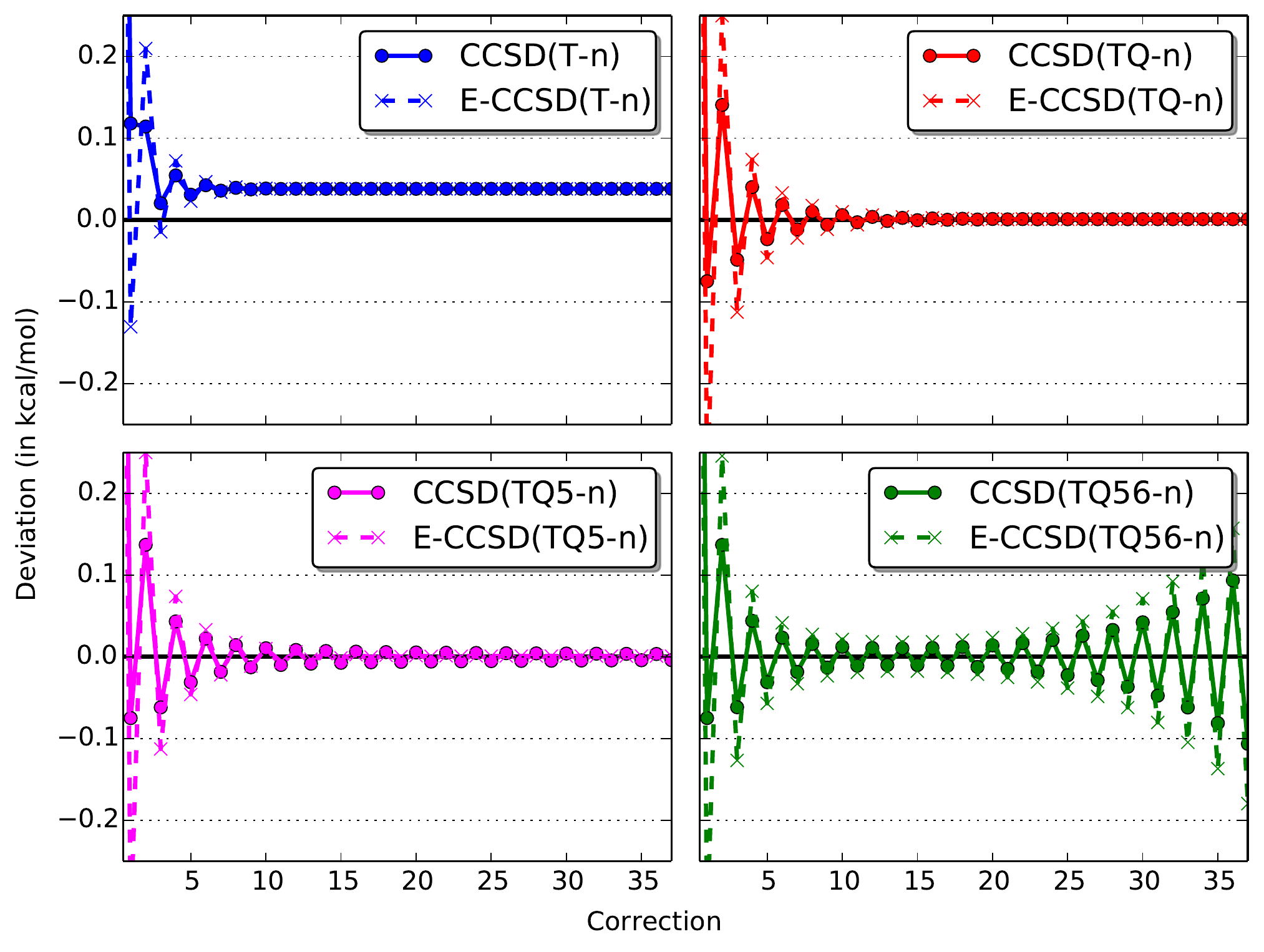}
   \caption{Total deviations (in kcal/mol) of the CCSD([$m_{\text{Q}}$]--$n$) and E-CCSD([$m_{\text{Q}}$]--$n$) series from the frozen-core/aug-cc-pVDZ FCI correlation energy for Ne.}
   \label{neon_cc_and_ecc_series_figure}
\end{figure}
As discussed in \ref{theory_order_corrections_section}, the CC[$m_{\text{P}}$]([$m_{\text{Q}}$]--$n$) and E-CC[$m_{\text{P}}$]([$m_{\text{Q}}$]--$n$) series will exhibit identical convergence radii, due to their common Jacobian in \ref{e_cc_jacobian}. However, as initially shown in Ref. \citenum{e_ccsd_tn_jcp_2016} through results for the lowest-order corrections of the CCSD(T--$n$) and E-CCSD(T--$n$) series, the rate of convergence towards the CC[$m_{\text{Q}}$] target energy will be different. In \ref{neon_cc_and_ecc_series_figure}, this is exemplified for Ne not only for the CCSD(T--$n$) and E-CCSD(T--$n$) series through higher orders, but also for differing choices of CC[$m_{\text{Q}}$] target states. From these results, we note how the oscillations through the lowest orders are considerably dampened in the CCSD([$m_{\text{Q}}$]--$n$) series over the corresponding E-CCSD([$m_{\text{Q}}$]--$n$) series, a direct consequence of the fact that the order expansions of the former family of series are markedly more balanced than those of the latter, cf. the discussions in \ref{theory_order_corrections_section} and, in particular, Ref. \citenum{e_ccsd_tn_jcp_2016}. Thus, in terms of practical convergence, the CC[$m_{\text{P}}$]([$m_{\text{Q}}$]--$n$) variants will always be preferable.

%
%
\subsection{Hydrogen fluoride with a stretched bond}\label{hf_dist_subsection}
\begin{figure}[!ht]
        \centering
        \includegraphics[scale=0.70,bb=7 9 578 432]{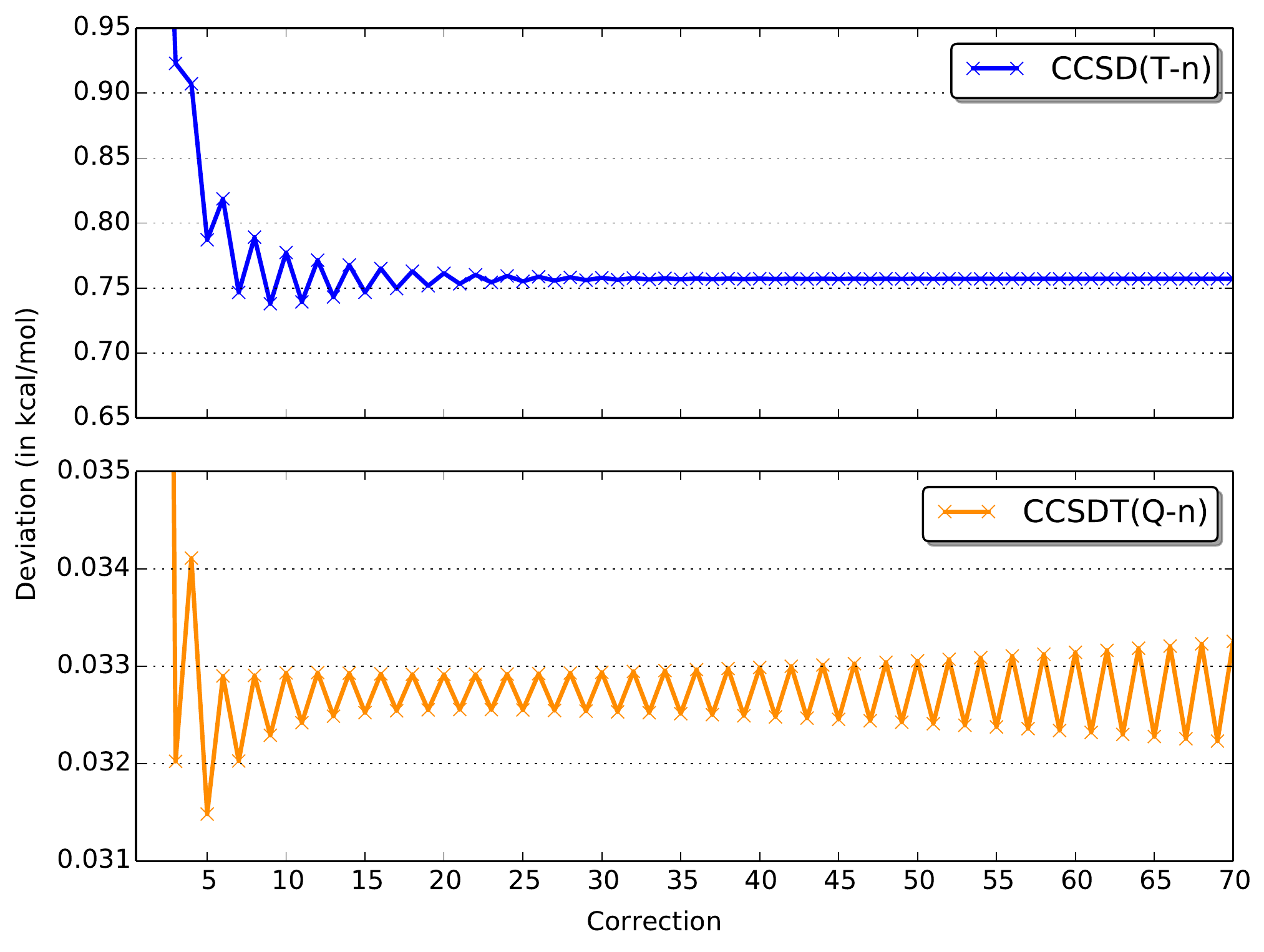}
   \caption{Total deviations (in kcal/mol) of the CCSD(T--$n$) and CCSDT(Q--$n$) series from the frozen-core/cc-pVDZ FCI correlation energy for stretched HF.}
   \label{hf_ccpvdz_figure}
\end{figure}
In Ref. \citenum{mp_divergence_olsen_jcp_2000}, the convergence of the MP series for hydrogen fluoride was investigated for combinations of two different geometries and two different basis sets. At the equilibrium geometry, the MP series was found to be rapidly convergent in the cc-pVDZ basis set and excessively oscillating (ultimately divergent) in the aug-cc-pVDZ basis. Upon distorting the geometry, however, the MP series was found to diverge in the standard cc-pVDZ basis as well, with an avoided crossing between the ground state and a back-door intruder well within the unit circle.
\begin{figure}[!ht]
        \centering
        \includegraphics[scale=0.70,bb=7 9 559 414]{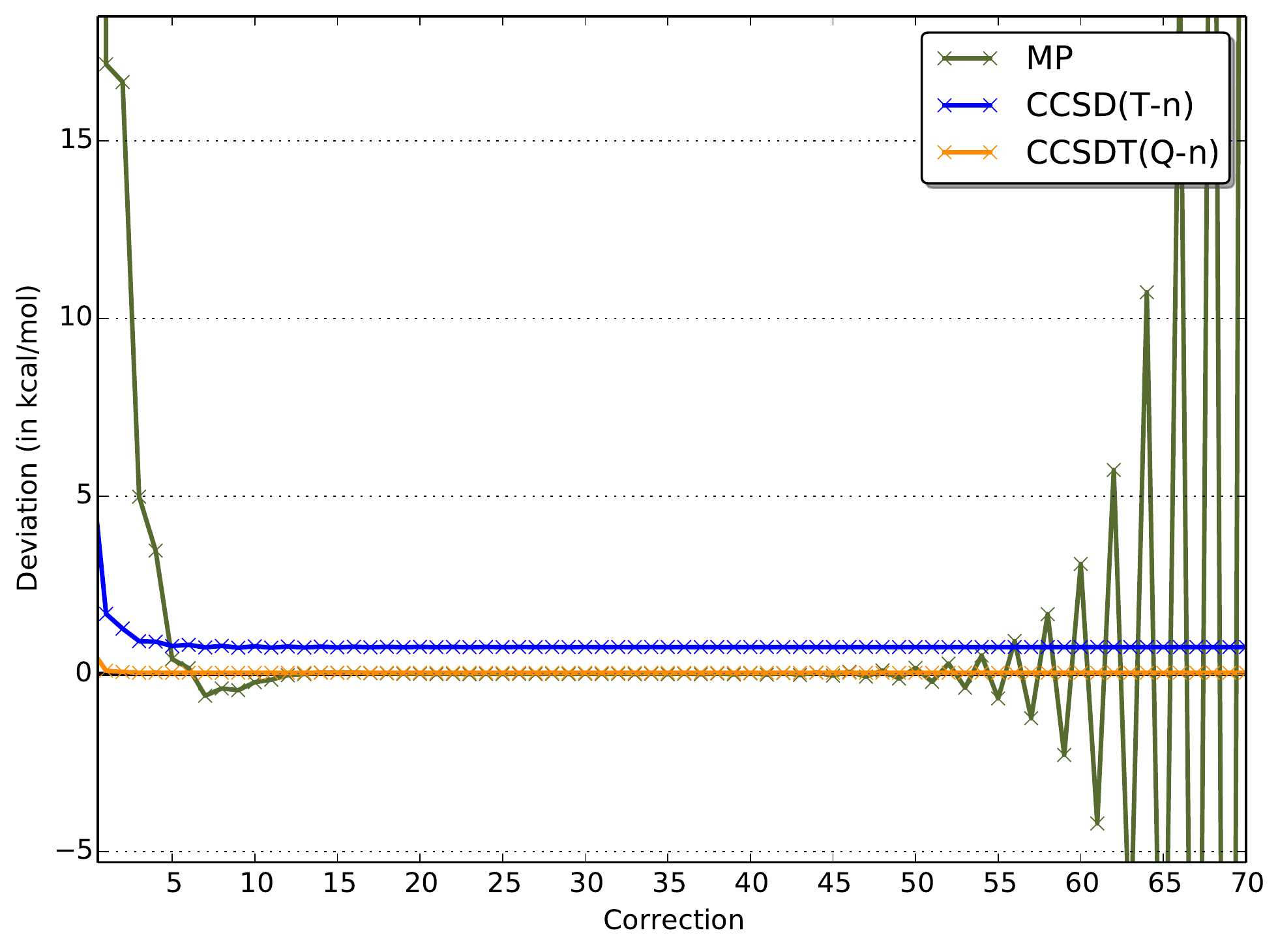}
   \caption{Total deviations (in kcal/mol) of the MP, CCSD(T--$n$), and CCSDT(Q--$n$) series from the frozen-core/cc-pVDZ FCI correlation energy for stretched HF.}
   \label{hf_ccpvdz_2_figure}
\end{figure}
As opposed to the intruder state in the case of the neon atom (\ref{neon_subsection}), the back-door intruder for distorted HF is not dominated by an exorbitant amount of high-level excitations, but rather by a mixture of double, triple, and, in particular, quadruple excitations. In fact, the quadruples contribution to the description of the intruder state is so pronounced that if these are left out of the cluster operators in the CC-based series, the position of the avoided crossing is observed to shift from just within the unit circle, as for the CCSD(TQ--$n$) series in \ref{convergence_table} and the CCSDT(Q--$n$) series in \ref{hf_ccpvdz_figure}, to a value slightly below $z=-1$ for the CCSD(T--$n$) series, also shown in \ref{hf_ccpvdz_figure}. Thus, whereas both HF- and CC-based expansions are in general divergent for distorted hydrogen fluoride, it is possible to form truncated expansions that do converge by choosing a target state that does not include quadruple (and higher-level) excitations. This conclusion is thus equivalent to the one for the neon atom in \ref{neon_divergence_subsubsection}, for which the omission of hextuple and higher-level excitations in the cluster operator leads to a convergent series. For both of the CC-based series in \ref{hf_ccpvdz_figure}, however, a \textit{practical} convergence towards the respective target energy (e.g., CCSDT or CCSDTQ) is observed, cf. \ref{hf_ccpvdz_2_figure}. Specifically, the divergence of the CCSDT(Q--$n$) series through the first 70 corrections is manifested only in very small energy oscillations of the order of 0.001 kcal/mol, which are negligible for all practical purposes and not even visible on the energy scale of \ref{hf_ccpvdz_2_figure}. 

%
%
\subsection{Singlet methylene}\label{methylene_subsection}
\begin{figure}[!ht]
        \centering
        \includegraphics[scale=0.70,bb=7 9 559 413]{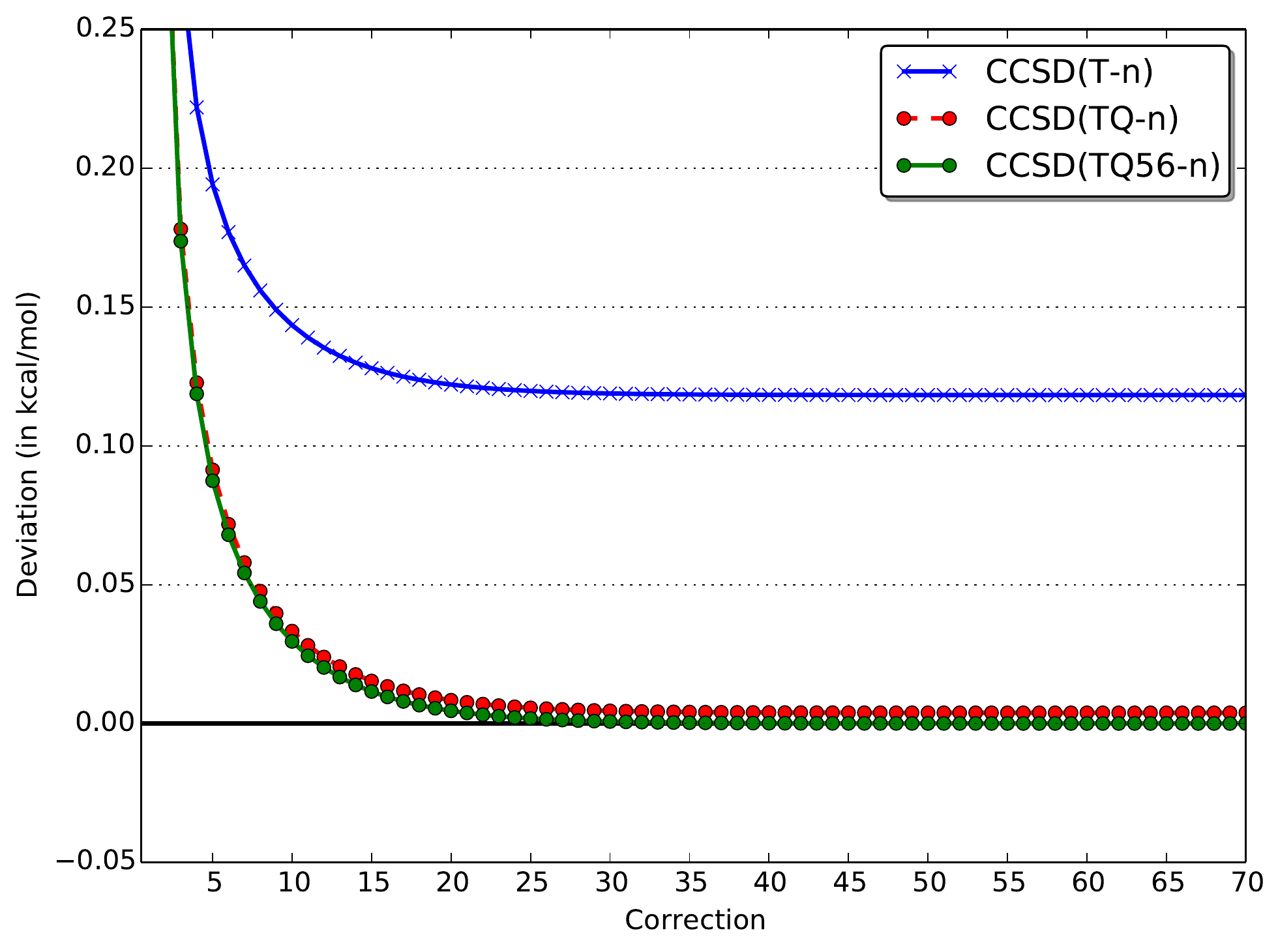}
   \caption{Total deviations (in kcal/mol) of the CCSD(T--$n$), CCSD(TQ--$n$), and CCSD(TQ56--$n$) series from the frozen-core/cc-pVDZ FCI correlation energy for CH$_2$.}
   \label{ch2_figure}
\end{figure}
Turning next to the case of singlet methylene, we note how the radius of convergence of the MP series was previously found in Ref. \citenum{mp_divergence_olsen_jcp_2000} to be larger than $1.0$ in the cc-pVDZ  basis (as well as in the aug-cc-pVDZ basis), thereby validating a convergent MP series. However, an avoided crossing between the ground state and a low-lying doubly excited state of the same spin and spatial symmetry was found at a positive $z$-value of about $1.2$ (that is, in the vicinity of, but still outside the boundary of the unit circle), which is confirmed for all of the tested perturbation series in \ref{convergence_table}. In particular, the position of this crossing is not observed to shift significantly with changes in the truncation level, [$m_{\text{Q}}$]. However, despite the fact that this excited state does not represent a front-door intruder state, the presence of the avoided crossing near the boundary of the unit circle is seen in \ref{ch2_figure} to cause a monotonic, but distinctly slow rate of convergence for all of the tested series, regardless of the choice of target state. In particular, we note how the shapes of the energy curves for singlet methylene in \ref{ch2_figure} are completely different from the ones for Ne and HF, cf. \ref{neon_figure} and \ref{hf_ccpvdz_figure}, differences which are intimately related to those of the energy profiles of front- and back-door intruders as described in, for instance, Ref. \citenum{mp_divergence_christiansen_cpl_1996}.

%
%
\subsection{The fluoride anion}\label{f_anion_subsection}

Finally, we close the present Section with the most challenging of the four examples studied here, namely the fluoride anion in the aug-cc-pVDZ basis set, for which the MP series was previously found in Refs. \citenum{mp_divergence_olsen_jcp_1996} and \citenum{mp_divergence_olsen_jcp_2000} to diverge from the very onset of the expansion (from the MP3 model and onwards), essentially exploding up through higher orders. Although F$^{-}$ is isoelectronic with Ne, and the weight of the HF determinant is approximately the same for the two species in the cc-pVDZ and aug-cc-pVDZ basis sets ($93\%-97 \%$), being an anion, the need for augmented basis sets in the accurate description of F$^{-}$ is of utmost importance. This is also recognized from the fact that the HOMO-LUMO gap in F$^{-}$ is considerably lowered upon augmenting the cc-pVDZ basis by diffuse functions, from $55$ eV in the cc-pVDZ basis to less than $20$ eV in the aug-cc-pVDZ basis. In Ref. \citenum{mp_divergence_olsen_jcp_2000}, an avoided crossing between the ground state and a relatively strongly coupled back-door intruder was observed at $\operatorname{Re}(\xi) \simeq -0.6$ for the MP series, and the position of the crossing closer to $z=0$---as well as the larger coupling between the two states---were used to explain the more rapid divergence for F$^{-}$ over that for Ne.

\begin{figure}[!ht]
        \centering
        \includegraphics[scale=0.70,bb=7 9 577 433]{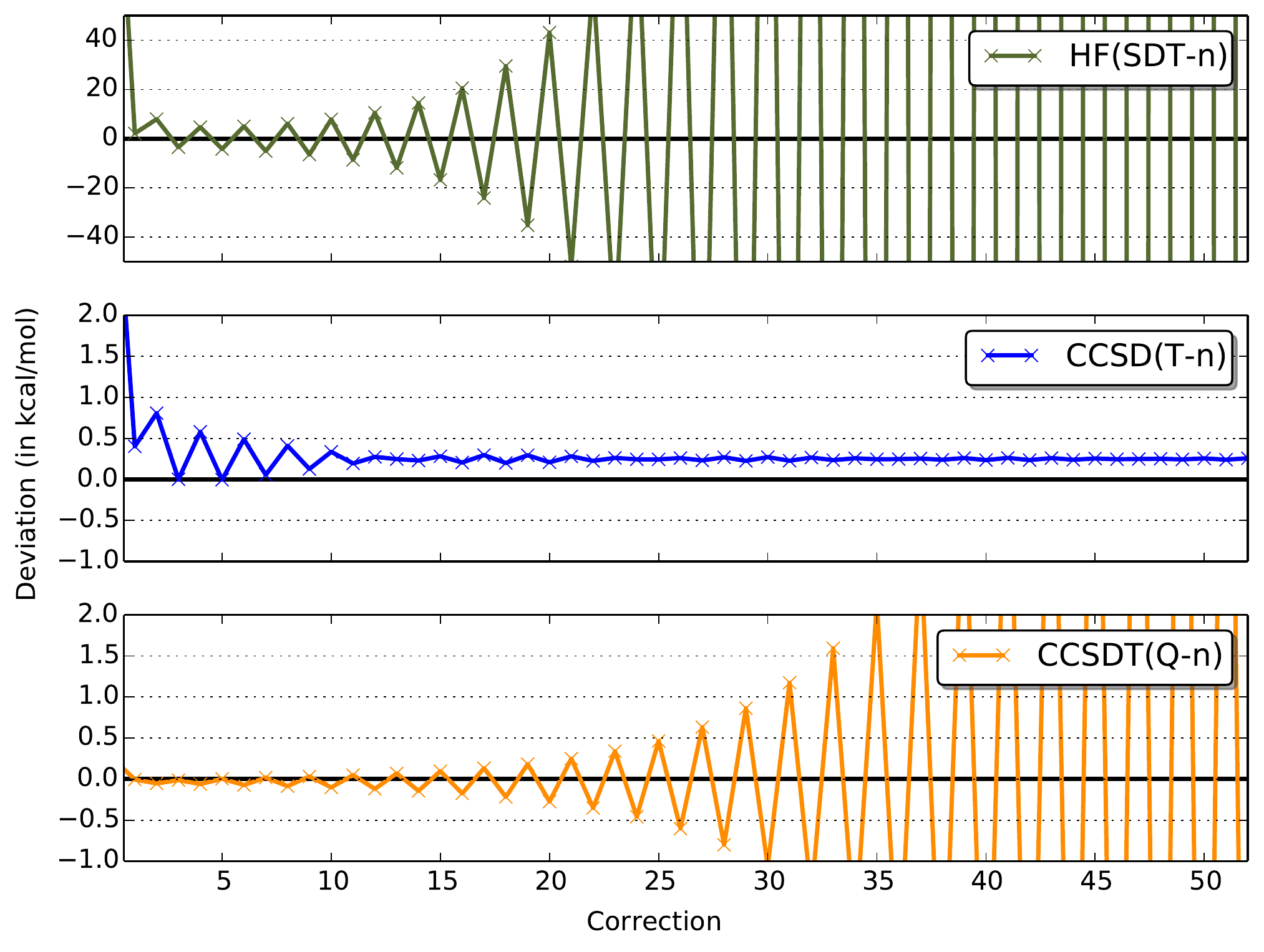}
   \caption{Total deviations (in kcal/mol) of the HF(SDT--$n$), CCSD(T--$n$), and CCSDT(Q--$n$) series from the frozen-core/aug-cc-pVDZ FCI correlation energy for F$^{-}$. Please note the difference in scales in the individual plots.}
   \label{f_anion_1_figure}
\end{figure}
\begin{figure}[!ht]
        \centering
        \includegraphics[scale=0.70,bb=7 9 558 413]{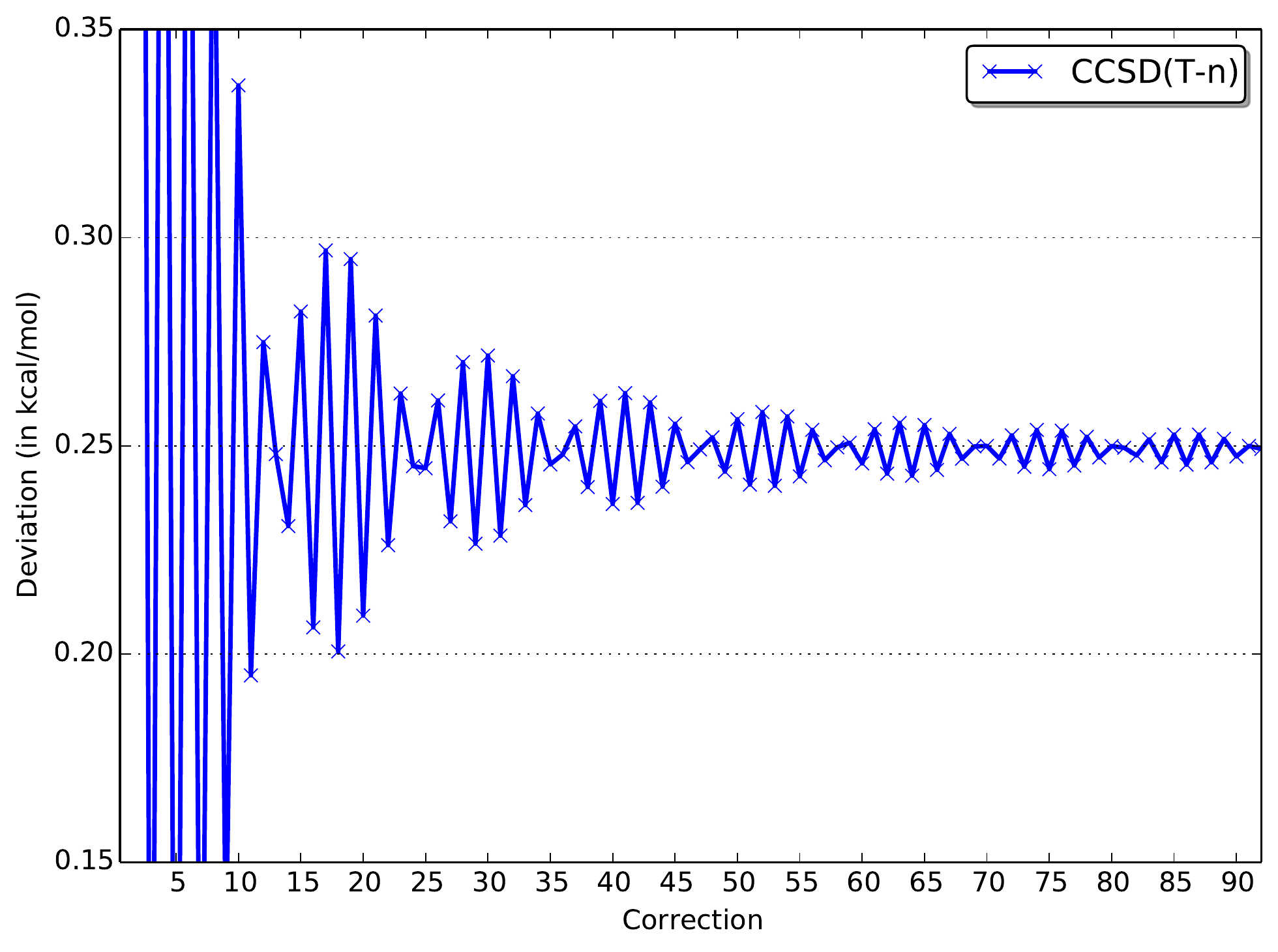}
   \caption{Detailed view on the total deviation (in kcal/mol) of the CCSD(T--$n$) series from the frozen-core/aug-cc-pVDZ FCI correlation energy for F$^{-}$.}
   \label{f_anion_2_figure}
\end{figure}

In analogy with the studies of the neon atom in \ref{neon_subsection} and distorted HF in \ref{hf_dist_subsection}, the wave function for the intruder state may be decomposed into contributions from individual excitation levels. By doing so, we find that more than $60 \%$ of the wave function consists of singly, doubly, and triply excited determinants. Furthermore, for an HF(SD--$n$) order expansion of the CCSD correlation energy, no back-door intruder is observed. Collectively, this indicates how the inclusion of triple excitations is the primary cause for the divergence of the HF(SDT--$n$) series, see \ref{f_anion_1_figure}. However, in contrast to this result, we note from \ref{f_anion_1_figure} how the CCSD(T--$n$) triples series converges. This difference between the HF(SDT--$n$) and CCSD(T--$n$) series hence indicates that the choice of parent state may also impact the convergence behavior of CC perturbation series, despite being a minor overall factor compared to the influence of basis set diffuseness and level of target state. Relating this back to the analysis in \ref{theory_scan_section}, this implies that if pronounced differences exist between the amplitudes of $T$ and $\tilde{T}$ (i.e., the difference in \ref{jacobian_difference} is substantial), this may too impact the radius of convergence. The actual convergence of the CCSD(T--$n$) series for F$^{-}$, though, is observed to be highly irregular and almost pulse-like, as may be realized from the detailed view of the order corrections in \ref{f_anion_2_figure}. 

For the sake of completeness, it should also be noted that the lowest Jacobian eigenvalue obtained during the CCSD(T--$n$) $z$-scan for F$^{-}$ contained a very small imaginary component for $z$-values in the interval $[-1.0;-0.9]$. However, based on the CCSD(T--$n$) energy corrections in \ref{f_anion_1_figure} and \ref{f_anion_2_figure} alongside the fact that this imaginary component was found to be a full six orders of magnitude smaller than the (positive) real component, we conclude that the CCSD(T--$n$) series for F$^{-}$ is indeed convergent. It should also be mentioned that the $z$-scan of the Jacobian for the CCSD(TQ--$n$) series was the only case in which the eigenvalue with the smallest real part had a large imaginary component. In this case, it is therefore not possible to use the $z$-scan on the real axis to determine whether the series is convergent or divergent. However, the perturbation expansion (not shown) was observed to diverge rapidly.

Of all the examples considered in the present work, we only encountered a single example---namely, the HF(SDT--$n$) and CCSD(T--$n$) series for F$^{-}$---for which the choice of parent state could be attributed as having an effect on the convergence behaviour, while the choice of target state is generally a much more important factor (see \ref{neon_subsection} and \ref{hf_dist_subsection}). This contrast in significance is also illustrated by the fact that the CCSD(TQ--$n$) and CCSDT(Q--$n$) series for F$^{-}$ both diverge, see \ref{convergence_table} and \ref{f_anion_1_figure}. It is, however, worth noting---by pragmatically comparing the corrections of the various series in \ref{f_anion_1_figure}---how the use of the lowest-order corrections of any of the CC-based series is not invalidated in the case of F$^{-}$, as opposed to the situation for HF-based analogues, for which only the second-order (MP2) correction is typically considered meaningful.

Now, the special case of the convergent CCSD(T--$n$) series aside, the results in \ref{f_anion_1_figure} highlight how the majority of the problems for F$^{-}$ originates from the underlying MP partitioning of the electronic Hamiltonian used in both the HF- and CC-based series, that is, the division of $\hat{H}$ into the zeroth-order Fock operator and a sizeable perturbation in the form of the fluctuation potential. This is so because the combination of an electron-rich ground state and an exceedingly diffuse excited state (archetypal for anions in augmented basis sets) makes for dominant contributions to either of the $\textbf{J}$ and $\tilde{\textbf{J}}$ Jacobians that induce avoided crossings between such states for negative values of $z$. This notion is further substantiated by probing for negative excitation energies using the $\textbf{J}^{\text{com}}$ Jacobian in \ref{jacobian_common}, for which the same crossing as in the HF(SDT--$n$) and CCSDT(Q--$n$) series is again observed. 

%
%
\section{Summary and conclusions}\label{summary_conclusion_section}

The radius of convergence of various HF- and CC-based MP perturbation expansions with different CC target states has been determined for a selection of prototypical closed-shell examples that are known from the literature to have slowly convergent or even divergent MP series. In particular, we have probed for the presence of potential intruder states in truncated MP series as well as in the recently proposed CCSD(T--$n$) triples and CCSD(TQ--$n$)/CCSDT(Q--$n$) quadruples series, in turn by probing for zero- or negative-valued excitation energies as calculated from the CC Jacobian for each of the series. The similarities between the intruder states encountered in HF- and CC-based perturbation theory have been interpreted in terms of structural similarities between the resulting Jacobians for the different series. By focusing on the common and dominating contribution to these, we have been able to confirm and supplement the main conclusions of the original MP convergence study in Ref. \citenum{mp_divergence_olsen_jcp_2000}. 

We have detailed how perturbation theory formulated around an MP partitioning of the electronic Hamiltonian, be that based on an HF or a CC ground state wave function, will be prone to back-door intruder states whenever a description of diffuse continuum states is possible, as, for instance, is the case when an augmented basis set is used or the target state contains highly excited determinants. Thus, in general, a divergent HF- or CC-based series for a given combination of molecular structure and basis set can likely be made convergent by reducing the excitation level of the target state and/or reducing the diffuseness of the basis set. On the other hand, the choice of parent state has typically (but not always) no effect on the convergence behaviour. 

Furthermore, we have reiterated the statement saying that back-door intruders are indeed prevalent for electron-rich systems with dense ground state wave functions. From a comparison of the Jacobians of HF- and CC-based perturbation theory, both of these conclusions have been traced back to the fundamental MP partitioning of the Hamiltonian, namely the fact that the perturbation within this framework, i.e., the fluctuation potential, is substantial in size, in contrast with the general premise of perturbation theory where the perturbation in itself is assumed small. Thus, we argue that the observed divergences are artefacts of the MP partitioning and, as such, inherent to any perturbation expansion that makes use of this. Perhaps somewhat counterintuitive, we have also illustrated how the magnitude of the actual energy difference, which is expanded in orders of the perturbation in any given perturbation series, is completely irrelevant for the actual convergence behaviour of said series. 

However, divergences aside, the CC-based perturbation series have all been observed to behave significantly more stably than the MP series or any of its truncated variants. Furthermore, we have found the bivariational CCSD(T--$n$) and CCSDT(Q--$n$) series to be capable of remedying the disordered oscillations at lower orders in the perturbation---which is traditionally one of the premier characteristics of divergent perturbation expansions---even in cases where these series formally diverge at higher orders. As such, despite the fact that the present analysis has been limited to minimal systems (atoms and diatomics), we find that even when potential divergences exist, as, for instance, in the case of the CCSDT(Q--$n$) series for stretched hydrogen fluoride and the fluoride anion, these are not found to invalidate the use of lower-order models such as, e.g., the CCSDT(Q--3) and CCSDT(Q--4) models. This is indeed an important point to stress, as formal divergences remain the norm rather than the exception for perturbation expansions in orders of the MP fluctuation potential, irrespective of whether these are based on an uncorrelated HF state or a correlated CC state. 

%
%
\section*{Acknowledgments}

J. J. E., K. K., and P. J. acknowledge support from the European Research Council under the European Union's Seventh Framework Programme (FP/2007-2013)/ERC Grant Agreement No. 291371, D. A. M. acknowledges support from the US National Science Foundation (NSF) under grant number ACI-1148125/1340293 and from the Arnold and Mabel Beckman Foundation as an Arnold O. Beckman postdoctoral fellow, and J. O. acknowledges support from the Danish Council for Independent Research, DFF-4181-00537.

\newpage

\providecommand*\mcitethebibliography{\thebibliography}
\csname @ifundefined\endcsname{endmcitethebibliography}
  {\let\endmcitethebibliography\endthebibliography}{}

\end{document}